\documentclass[a4paper,11pt]{article}
\pdfoutput=1 % if your are submitting a pdflatex (i.e. if you have
\usepackage{jcappub} % for details on the use of the package, please
\usepackage[T1]{fontenc} % if needed
\usepackage{graphicx}
\usepackage{longtable}
\usepackage{float}
\usepackage{dcolumn}
\usepackage{bm}
\usepackage{appendix}
\usepackage{multirow}
\usepackage{color}
\usepackage{soul}
\usepackage[normalem]{ulem}
\usepackage{amsfonts}
\usepackage{amsmath}
\usepackage{amssymb}
\usepackage{hyperref}
\usepackage{subfigure}

\title{Observational Constraints on Warm Inflation in Loop Quantum Cosmology}

\author[a,b]{Micol Benetti,} 
\affiliation[a]{Dipartimento di Fisica  ``E. Pancini", Universit\`a di Napoli  
``Federico II", Via Cinthia, I-80126, Napoli, Italy}
\affiliation[b]{Istituto Nazionale di Fisica Nucleare (INFN), sez. di Napoli, 
Via Cinthia 9, I-80126 Napoli, Italy}

\author[c]{L. L. Graef} 
\affiliation[c]{Instituto de F\'{\i}sica, Universidade Federal Fluminense, 
Avenida General Milton Tavares de Souza s/n, Gragoat\'a, 24210-346 Niter\'oi, 
Rio de Janeiro, Brazil}

\author[d]{and Rudnei O. Ramos} 
\affiliation[d]{Departamento de F\'{\i}sica Te\'orica, Universidade do Estado do 
Rio de Janeiro, 20550-013 Rio de Janeiro, RJ, Brazil}

\emailAdd{micol.benetti@na.infn.it}
\emailAdd{leilagraef@if.uff.br}
\emailAdd{rudnei@uerj.br}

\abstract{

By incorporating quantum aspects of gravity,  Loop Quantum
Cosmology (LQC) provides a self-consistent extension of the
inflationary scenario, allowing for modifications in the primordial
inflationary power spectrum with respect to the standard General
Relativity one. We investigate such modifications and explore the
constraints imposed by the Cosmic Microwave Background (CMB) Planck
Collaboration data on the Warm Inflation (WI) scenario in the LQC
context.  We obtain useful
relations between the dissipative parameter of WI and the bounce scale
parameter of LQC.   
We also find that the number of required e-folds of expansion
from the bounce instant till the moment the observable scales crossed
the Hubble radius during inflation can be smaller in WI than in CI. In particular,
we find that this depends on how large is the dissipation in WI,
with the amount of required e-folds decreasing with the increasing of the
dissipation value. {}Furthermore, by performing a Monte Carlo Markov Chain analysis 
for the considered WI models, we find good  agreement of the model with the data. 
This shows that the WI models studied here can explain the current observations 
also in the context of LQC.

}
\begin{document}
\maketitle
\flushbottom

%%%%%%%%%%%%%%%%%%%%%%%%%%%%%%%%%%%%%%%%%%%%%%%%%%%%%%%%%%%%%%%%%%%%%%%%%%
\section{Introduction}
\label{intro}

In the last years we have witnessed the release of a large
amount of precision data, from the Cosmic  Microwave Background
(CMB)~\cite{Aghanim:2015xee,Bennett:2012zja} to large
scale structures~\cite{SDSS,Paris:2017xme,Scolnic:2017caz}, 
including Barionic Acoustic Oscillation data~\cite{Dawson:2012va,Bautista:2017wwp}, 
both strong and weak lensing~\cite{Suyu:2016qxx,Abbott:2017wau,Martinet:2019abk}, 
galaxy cluster number counts~\cite{Ade:2015gva,DeBernardis:2016pdv} and so on, 
up to gravitational waves detections~\cite{Abbott:2016blz,TheLIGOScientific:2017qsa,Monitor:2017mdv}.  
This has allowed us to obtain valuable information about the nature 
and the evolution of the universe, as well as the mechanisms operating 
at the very early times (see, e.g., refs.~\cite{SZ} and \cite{peebles}).  
The current paradigm for the cosmology of the early universe is inflation,
which besides of solving the problems of the standard Big Bang cosmology, provides a
causal explanation for the origin of the CMB anisotropies and the
large-scale structure of the universe~\cite{mukhanov} (see also
ref.~\cite{press}). The inflationary scenario was developed long
before accurate data were available, which makes it a very predictive
scenario. However, the recent CMB data imposed a big challenge for some
classes of inflationary models by putting severe constrains on many
of them~\cite{Akrami:2018odb}.

The description of inflation can be classified in two scenarios
according to the dynamics of the inflaton field. In the Cold Inflation
(CI) scenario, interactions of the inflaton with other field degrees
of freedom are not enough to counter balance the dilution of any
possible pre-existing or newly formed radiation and the universe super
freezes. Density perturbations are originated as quantum fluctuations
of the inflaton field~\cite{lyth}.  In the Warm Inflation (WI)
scenario~\cite{Berera:1995ie} (see also refs.~\cite{Berera:2008ar,BasteroGil:2009ec} for
reviews), on the other hand, the interactions of the inflaton with
other field degrees of freedom (and also among the latter) can be
sufficient to produce a quasi-stationary thermalized radiation bath
throughout inflation. The primary source of density fluctuations in
this case can come entirely  from thermal fluctuations originated in
the radiation bath and transported to the inflaton field as adiabatic
curvature
perturbations~\cite{taylor_berera,Hall:2003zp,Graham:2009bf,BasteroGil:2011xd,Bastero-Gil:2014jsa}.
In the WI picture, the presence of nontrivial dynamics, accounting for
dissipative and  stochastic effects, cause a significant  impact on
the usual observational quantities like in the tensor-to-scalar ratio,
$r$, the spectral index, $n_{s}$, and the non-Gaussianity parameter,
$f_{NL}$~\cite{Bartrum:2013fia,Bastero-Gil:2014oga,Bastero-Gil:2014raa,Visinelli:2014qla,Ramos:2013nsa,
  Benetti:2016jhf}. Due to these modifications, some classes of
inflaton potentials excluded in the CI context by the data can be
rehabilitated in the WI context, as it is the case of the monomial
chaotic potentials for instance.  In the WI scenario the coupling between the
inflaton and other fields might be strong enough to lead to a
significant radiation production rate, while still preserving the
expected  flatness of the inflaton potential. The radiation production
during WI can compensate for the  supercooling of the universe observed in
CI, thus making possible for a smooth transition from the inflationary
accelerated expansion to the radiation dominate phase, without the
need or a presence of a (pre)reheating phase following the end of the
inflationary regime. 

One important aspect to point out is that the WI scenario, as is also
true for the CI one, they are both sensitive to the ultraviolet (UV)
physics, and their successes are tightly dependent on the
understanding of such UV physics in general.  Inflationary space-times
inherit the big-bang singularity. Physically, this occurs because one
continues to use General Relativity (GR) theory even in the Planck
regime where it is supposed not to be applicable. It is widely
expected that new physics in this regime will resolve the singularity,
significantly changing the very early history of the universe.  One of
the possible scenarios that takes into account a new physics in this
high energy regime  is Loop Quantum Gravity (LQG), which is believed
to be a possible candidate for a quantum theory of gravity (see, e.g.,
refs.~\cite{LQC23,LQC24,LQC25,LQC26,LQC27} for some recent
reviews). Loop Quantum Cosmology (LQC) arises as the result of
applying principles of LQG to cosmological settings (see, e.g.,
refs.~\cite{LQC25,LQC26,LQC27} for recent reviews). In LQC the quantum
geometry creates a brand new repulsive force, which is totally
negligible at low space-time curvature but rises very rapidly in the
Planck regime, overwhelming the classical gravitational attraction. In
cosmological models, Einstein's equations undergo  modifications in
the Planck regime. {}For matter satisfying the usual energy
conditions, any time a curvature invariant grows to the Planck scale,
quantum geometry effects dilute it, resolving singularities of the
GR~\cite{LQC28,LQC29,LQC30}. These quantum gravitational effects are
expected to dominate the Planck era of the universe causing a quantum
bounce  to appear and that replaces the classical big bang
singularity.  Usually, in these scenarios, where inflation is studied in the context of
LQC, the inflaton potential is assumed to be simply a quadratic potential for the inflaton field. Very close to the bounce quantum gravity
effects dominate the dynamics, but these effects lessen shortly after
such that the potential $V(\phi)$ begins to prevail, starting the inflationary phase. Though this picture is the most common in the literature
and the one we will adopt also here, there are also other possibilities, with the dominant field at the bounce depending on the cosmological
scenario being considered. Also, in the context of inflation it is not
necessary to only consider scalar fields with a quadratic potential, with
other potentials being equally possible. In this work, we will consider
in fact a quartic monomial chaotic potential for the inflaton. Therefore, LQC
provides an interesting arena to incorporate the highest energy
density and curvature stages of the universe into cosmological models,
where questions about Planck-scale physics and initial conditions for
inflation can be addressed~\cite{agullo_morris}.

Cosmological perturbations are generally described by quantum fields
on classical, curved space-times.  However this strategy  cannot be
trivially justified in the quantum gravity era, when  curvature and
matter densities are of Planck scale.  Nevertheless, using techniques
from LQG, the standard theory of cosmological perturbations was
extended to overcome this limitation~\cite{16pp}. The dressed metric
approach~\cite{13pp,14pp,15pp,16pp} is able to provide this extension,
while having  the advantage of allowing ones to describe the main
perturbation equations in a form analogous to the classical one~\cite{16pp,15pp}. 
Also, the pre-inflationary evolution makes scalar and tensor
perturbations reach the onset of inflation in an excited
state~\cite{14pp}, so that the primordial spectra that source the CMB
anisotropies acquire extra features with quantum gravitational origin
from the pre-inflationary era.

Previous studies on both CI and WI in the context of LQC showed interesting 
features regarding properties of the pre-inflationary phase and the start of inflation 
itself~\cite{zhu,Herrera:2010yg,39wilqc,40wilqc,41wilqc,42wilqc,43wilqc,44wilqc,leila_rudnei,Bedic:2018gqu}.
In the present work we focus for the first time on the problem of confronting
the observational predictions of WI in LQC with the latest CMB
data. By accounting for the modifications of the spectrum of
primordial perturbations resulted from the LQC, we compute the number
of extra e-folds required in this model for it to be compatible with
the observations and put constraints on the LQC parameters when in the
WI context. {}Finally, we focus on analyzing the relation of these
characteristic parameters of LQC  with the dissipative parameter of WI.

This paper is organized as follows.  In section~\ref{TC} we describe
the theoretical context of the  WI scenario in LQC, deriving the equations 
we use in our analysis. The method and observational dataset are discussed
in section~\ref{analysis}, where we also present and discuss our
results. {}Finally, in section~\ref{conclusion} we summarize our conclusions.

%%%%%%%%%%%%%%%%%%%%%%%%%%%%%%%%%%%%%%%%%%%%%%%%%%%%%%%%%%
\section{Theoretical context}
\label{TC}

In this section we briefly review some of the most relevant aspects of
LQC and WI, then showing the expressions for their observables in the
WI scenario in LQC, which will be considered in our analysis in the next section.

%%%%%%%%%%%%%%%%%%%%%%%%%%%%%%%%%%%%%%%%%%%%%%%%%%%%%%%%%
\subsection{Loop Quantum Cosmology dynamics}

 The spatial geometry in LQC is encoded in the volume of a fixed
 fiducial cubic cell, rather than the scale factor $a$, and is given
 by 
\begin{equation}
v = \frac{4 {\cal V}_0 a^{3} M_{\rm Pl}^2}{\gamma},
\label{eqv}
\end{equation}
where ${\cal V}_0$ is the comoving volume of the fiducial cell,
$\gamma$ is the Barbero-Immirzi parameter of LQC, whose numerical
value we set as given by $\gamma\simeq 0.2375$~\cite{Meissner:2004ju},
and $M_{\rm Pl}\equiv 1/\sqrt{8 \pi G} = 2.4 \times 10^{18}$GeV is the
reduced Planck mass. The conjugate momentum to $v$ is denoted by $b$
and it is given by $b=-\gamma P_{(a)}/(6 a^2 {\cal V}_0 M_{\rm
  Pl}^2)$,  where $P_{(a)}$ is the conjugate momentum to the scale
factor.  

The solution of the LQC effective equations implies that the Hubble
parameter, $H$, can be written as
\begin{equation}
H=\frac{1}{2 \gamma \lambda} \sin(2 \lambda b).
\label{eqH}
\end{equation}
where $\lambda =\sqrt{\sqrt{3} \gamma/(2 M_{\rm Pl}^2)}$ and $b$
ranges over $(0, \pi/\lambda)$.   The energy density, $\rho$, relates
to the LQC variable $b$ through $\rho = 3 M_{\rm Pl}^2 \sin^2(\lambda
b)/(\gamma^2 \lambda^2)$.  Then, the {}Friedmann's equation in LQC
assumes the form~\cite{Ashtekar:2011rm},
\begin{equation}
\frac{1}{9}\left(\frac{\dot{v}}{v}\right)^{2} \equiv H^2 = \frac{1}{3
  M_{\rm Pl}^2} \rho \left(1- \frac{\rho}{\rho_{\rm cr}} \right),
\label{Hubble}
\end{equation}
where $\rho_{\rm cr} = 3 M_{\rm Pl}^2/(\gamma^2 \lambda^2)$. {}For
$\rho \ll \rho_{\rm cr}$ we recover GR as expected. The above
expression holds independently of the particular characteristics of
the inflationary regime. We can see from eq.~(\ref{Hubble}) that
the singularity is replaced by a quantum bounce for $H=0$, when the
density reaches the critical value $\rho_{\rm cr}$.  

In addition to the modifications at the background level from LQC, at
the perturbative level some of the relevant modes have physical
wavelengths comparable to the curvature radius at the bounce time.
Unlike what happens in the CI scenarios in GR, where it is usually
assumed that the  pre-inflationary dynamics does not have any effect
on modes observable in the CMB, in LQC the situation is
different. Using techniques from LQG, the standard theory of
cosmological perturbations can be extended to encompass the quantum
gravity regime, allowing to describe the main perturbation equations
in a form analogous to the classical one~\cite{16pp}.  Also, modes
that experience curvature are excited~\cite{parker}, i.e., large
wavelength modes are excited in the Planck regime that follows the
bounce. The main effect at the onset of inflation is that the quantum
state of perturbations is populated by excitations of these modes over
the Bunch-Davis vacuum, changing the initial conditions for
perturbations at the onset of inflation~\cite{14pp}.  As a
consequence, the scalar curvature power spectrum in LQC gets modified
with respect to GR, such that it can be written as (see, e.g.,
ref.~\cite{zhu} for more details and for a complete derivation)
\begin{equation}\label{PR1}
\Delta_\mathcal{R}(k)= |\alpha_{k} +\beta_{k}|^{2}
\Delta_\mathcal{R}^{GR}(k).
\end{equation}
with $\alpha_{k}$ and $\beta_{k}$ are the Bogoliubov coefficients
(where the pre-inflationary effects are codified) and
$\Delta_\mathcal{R}^{GR}$ is the GR form for the power spectrum. 
 It should be noticed that in the Bunch-Davis vacuum, 
as in the general relativity case, we have that the  Bogoliubov coefficients
in eq.~(\ref{PR1}) should reduce simply to~\cite{parker} 
\begin{equation}
\alpha_k \to \alpha_k^{\rm BD}=1, \;\;\; \beta_k \to\beta_k^{\rm BD} =0,
\label{alphabetaBD}
\end{equation}
which can be seen as no excited (or produced) particles in the vacuum.
In LQC, the change of the spectrum can be seen exactly as a result of the
change of the vacuum state with respected to the GR case. This change in the
vacuum state can be interpreted as a consequence of gravitational
particle production due to the bounce. This becomes even clearer when writing
the term $|\alpha_{k} +\beta_{k}|^{2}$ in eq.~(\ref{PR1}) as
\begin{equation}
|\alpha_{k} +\beta_{k}|^{2} = 1+ 2 |\beta_k|^2 + 2 {\rm Re}(\alpha_k \beta_k^*),
\label{alphabeta}
\end{equation}
where $|\beta_k|^2 \equiv n_k$ is associated with the number of excitations
in the mode $k$~\cite{14pp}.
By following the same notation used by the authors of ref.~\cite{zhu}, the above eq.~(\ref{PR1}) 
is parameterized as
\begin{equation}\label{PR}
\Delta_\mathcal{R}(k)= (1+\delta_{PL})\Delta_\mathcal{R}^{GR}(k).
\end{equation}
where the factor $\delta_{PL}$ is scale ($k$-)dependent and takes into
account the LQC corrections.  It is explicitly given by~\cite{zhu}
\begin{eqnarray}
\label{complete}
\delta_{PL} &=&  \left[1+\cos\left(\frac{\pi}{\sqrt{3}}\right)\right]
      {\rm csch}^{2}  \left(\frac{\pi k}{\sqrt{6} k_{B}}\right)
      \nonumber \\  &+& \sqrt{2} \sqrt{\cosh\left(\frac{2\pi
          k}{\sqrt{6}
          k_{B}}\right)+\cos\left(\frac{\pi}{\sqrt{3}}\right)}
      \cos\left(\frac{\pi}{2\sqrt{3}}\right) \nonumber \\ & \times &
               {\rm csch}^{2}\left(\frac{\pi
                 k}{\sqrt{6}k_{B}}\right)\cos(2k
               \eta_{B}+\varphi_{k}),
\end{eqnarray}
where
\begin{equation}\label{short}
\varphi_{k} \equiv \arctan \left\{\frac{{\rm
    Im}[\Gamma(a_{1})\Gamma(a_{2})\Gamma^{2}(a_{3} - a_{1} - a_{2})]}
       {{\rm Re}[\Gamma(a_{1})\Gamma(a_{2})\Gamma^{2}(a_{3} - a_{1} -
           a_{2})]}\right\},
\end{equation}
with $a_1,\,a_2,\, a_3$ defined as $a_{1,2} = (1\pm 1/\sqrt{3})/2 -i
k/(\sqrt{6} k_B)$ and $a_3=1-ik/(\sqrt{6} k_B)$ and the index $B$ in
the quantities  indicates that they are calculated at the bounce.  In
particular, $\eta_B$ is the conformal time at the bounce and $k_B=
\sqrt{\rho_c} a_B/M_{\rm Pl}$ is a characteristic  scale also at the
bounce, which is the shortest scale (or largest wave number $k$) that feels the
space-time curvature during the bounce. We should mention that the expression (\ref{PR})
is valid also in the presence of a small amount of radiation during the
bounce (which dillutes very fast) in addition to the dominant contribution from the inflaton's energy density. 
Comparing eq.~(\ref{complete}) with eq.~(\ref{alphabeta}) we identify
\begin{eqnarray}
2 |\beta_k|^2 &=& \left[1+\cos\left(\frac{\pi}{\sqrt{3}}\right)\right]
 {\rm csch}^{2}  \left(\frac{\pi k}{\sqrt{6} k_{B}}\right),
\\
2 {\rm Re}(\alpha_k \beta_k^*) &=& 
\sqrt{2} \sqrt{\cosh\left(\frac{2\pi
          k}{\sqrt{6}
          k_{B}}\right)+\cos\left(\frac{\pi}{\sqrt{3}}\right)}
      \cos\left(\frac{\pi}{2\sqrt{3}}\right) \nonumber \\ & \times &
               {\rm csch}^{2}\left(\frac{\pi
                 k}{\sqrt{6}k_{B}}\right)\cos(2k
               \eta_{B}+\varphi_{k}).
               \label{complete2}
\end{eqnarray}
The term $\cos(2k \eta_{B}+\varphi_{k})$ in eq.~(\ref{complete2})  
oscillates very fast,  so it has negligible effect when averaging out
in time.  Then, for any practical purpose, in observable quantities
the factor $\delta_{PL}$ can be simply considered as being given by
\begin{equation} \label{delta}
\delta_{PL} = \left[1+\cos\left(\frac{\pi}{\sqrt{3}}\right)\right]
      {\rm csch}^{2}\left(\frac{\pi k}{\sqrt{6}k_{B}}\right).
\end{equation}

Note that in this case $\delta_{PL}$ can simply be identified with 
$2n_k$, i.e., with the own number of excitations
in the mode $k$ which appears as a consequence of the quantum bounce in LQC.
Since this pre-factor represents the effects of the pre-inflationary
dynamics, it will appear similarly also in the power spectrum of WI models.
The difference, as we will see in the next section, is that in WI there are also
the effect of explicitly particle production due to the microscopic processes generating the
radiation bath. As a consequence,
the factor $2 n_k$ changing the spectrum due to the departure from the Bunch-Davis
vacuum will have two contributions: one, considered in the equations above, due to the LQC bounce effect (hereafter $2n_k^{\rm LQC}$) which is given by 
eq.~(\ref{delta}), and another due to the particle production inherent of the
WI dynamics which we are going to denote by  $2n_k^{\rm WI}$ (see. e.g., ref.~\cite{Ramos:2013nsa} for an explicit derivation
of the later).

The same way that the quantum bounce changes the scalar power spectrum like
eq.~(\ref{PR}), likewise the tensor spectrum in the LQC is modified as~\cite{zhu}
\begin{equation}
\label{PH}
\Delta_{T}(k)= (1+\delta_{PL})\Delta_{T}^{GR}(k),
\end{equation}
where $\Delta_{T}^{GR}(k)$ is the tensor spectrum in GR.
Again, the term in front of the GR result in eq.~(\ref{PH}) can be exactly
associated with the change of the Bunch-Davis vacuum due to the production
of excitations with mode $k$ out of the vacuum and
$1+ \delta_{PL} \equiv 1 +2n_k^{\rm LQC}$.

In the following we will denote the number of e-folds for the relevant scales at Hubble
crossing as $N_{*} \equiv ln(a_{\rm end}/a_{*}) \approx 60$, where
$a_{\rm end}$ is the scale factor at the end of inflation and all
quantities with subscript $*$ are evaluated when the mode crosses the horizon at 
$k_*=a_* H_*$. In LQC, in addition to the usual number of e-folds
at Hubble crossing $N_*$, it is necessary an extra amount of e-folds,
$\delta N$, in order for the predictions from the model to be
consistent with observations~\cite{zhu}. This is so because, as seen
from eq.~(\ref{PR}), after the effects of the pre-inflationary
dynamics from LQC are taken into account, the power spectra  are
generically scale-dependent through the correction $\delta_{PL}$,
eq.~(\ref{delta}),  and also exhibits oscillatory
features\footnote{Note that oscillatory features in the power spectrum
  in bouncing models is a generic result~\cite{Brandenberger:2017pjz}.}.   
  As a consequence, in CI in order to be consistent with observations, the universe must have
expanded at least~\cite{zhu} around $21$ e-folds from the bounce till Hubble
radius crossing of the observables scales, such as to allow for these
scale-dependent features to get sufficiently diluted away and not
spoiling the perturbation spectra of CMB (see also
ref.~\cite{Wilson-Ewing:2016yan} and references therein for a
discussion about these and other LQC effects). 

The total number of e-folds of expansion from the moment of the bounce
till today, $N_{\rm tot}$, is related to the LQC parameter $k_{B}$
through the equation~\cite{zhu},
\begin{equation}\label{kbN}
\frac{k_{B}}{a_{0}}=\sqrt{\frac{\gamma_{B}}{3}}\frac{a_{B}}{a_{0}}
m_{pl}=\sqrt{\frac{\gamma_{B}}{3}}m_{pl} e^{N_{\rm tot}}.
\end{equation}
We note that, by assuming an upper bound on $k_{B}$ as set by
eq.~(\ref{kbN}), it can be translated into constraints on the total
number of e-folds. This, in turn, leads to a lower bound on the extra
number of e-folds of inflation required in LQC, since $\delta N >
N_{\rm tot} - N_{*} - N_{\rm after}$, where $N_{\rm after} \equiv
ln(a_{0}/a_{\rm end})$.     
 We are interested in finding an upper bound value
for the scale $k_{B}$, which corresponds to a lower value for the number
of e-folds. The LQC effects considered in this paper only become important at
sufficiently small scales, namely for modes that feel the space-time curvature during the bounce. If there are very few e-folds of inflation, then
LQC predicts large departures from scale-invariance in the CMB, and is
ruled out.  Since for inflation lasting a very large number of e-folds,
$N_{\rm infl} \gg N_*$, the pre-inflationary effects
on CMB are completly diluted, then we are most interested in the sweet spot
of there being just enough inflationary e-folds so that there may be some
LQC effects, but that are not too strong to be ruled out. So, in the following we will hence be
assuming that the total number of e-folds of inflation is around $N_{*}$, i.e.,
$N_{infl} \approx N_{*}$. Therefore, as usual, we consider $N_{*}\approx 60 \approx
N_{\rm after}$, such that $\delta N > N_{\rm tot} - 120$.

In this work, we analyze the $\delta N$ value required in the WI models
in LQC. {}For this, we use the constrains on $k_{B}$ for WI in LQC
together with  eq.~(\ref{kbN}) above. We also compare the results
for $\delta N$ in CI and WI, when both are constructed in the LQC
context. Let us firstly review the scenario of WI in what follows.

%%%%%%%%%%%%%%%%%%%%%%%%%%%%%%%%%%%%%%%%%%%%%%%%%%%%%%%%%
\subsection{The warm inflation scenario}
\label{WIcases}

In WI dynamics the presence of radiation plays an important
role. Therefore, we must take into   account explicitly this component
such that the total energy density in eq.~(\ref{Hubble}) is given by 
\begin{equation}
\rho = \frac{{\dot \phi}^2}{2} + V(\phi) + \rho_R.
\label{rho}
\end{equation}
with the inflaton field, $\phi$, and the radiation energy density,
$\rho_R$.  In this work we consider the monomial quartic chaotic
potential for the inflaton,
\begin{equation}
V(\phi) = \frac{\Lambda}{4} \left(\frac{\phi}{M_{\rm Pl}}\right)^4,
\label{Vphi}
\end{equation}
where $\Lambda/M_{\rm Pl}^4$ denotes here the (dimensionless) quartic
coupling constant.  The background evolution equations for the
inflaton and for the radiation energy density, are given,
respectively, by
\begin{eqnarray}
&& \ddot \phi + 3 H \dot \phi + \Upsilon(\phi, T) \dot \phi +
  V_{,\phi}=0,
\label{eqphi}
\\ && \dot \rho_R + 4 H \rho_R = \Upsilon(\phi, T) \dot \phi^2,
\label{eqrhoR}
\end{eqnarray}
where $\Upsilon(\phi, T)$ is the dissipation coefficient in WI, which
can be a function of the temperature and/or the background inflaton
field. The dissipation coefficient embodies the microscopic physics
involved in the interactions between the inflaton and  the other
fields (and also among these), accounting for the non-equilibrium
dissipative processes arising from these
interactions~\cite{Berera:2008ar,BasteroGil:2010pb}.  {}For a
radiation bath of relativistic particles, the radiation energy density
is given by  $\rho_R=\pi^2 g_* T^4/30$, where $g_*$ is the effective
number of light degrees of freedom  ($g_*$ is fixed according to  the
dissipation regime and interactions form used in WI).

In our work we consider two different dissipation regimes, namely with
the dissipation coefficient showing a cubic and linear dependence with
the temperature of the thermal bath, which represent the most common
functional dependences derived from previous model building in
WI. {}For instance, the dynamics leading to the dissipation
coefficient with a cubic form emerges in the low temperature regime of
WI, in which the inflaton is coupled to heavy intermediate fields, and
those are in turn coupled to the light radiation fields. The decay of
the heavy intermediate fields into the light radiation bath fields
produces a dissipation coefficient with a cubic dependence on the
temperature of the thermal radiation bath produced, such that the
resulting dissipation coefficient can be well described by the
expression~\cite{Berera:2008ar,BasteroGil:2010pb,BasteroGil:2012cm}
\begin{equation}\label{upsilonT3}
\Upsilon_{\rm cubic} = C_{\rm cubic} \frac{T^3}{\phi^2}, 
\end{equation}
where $C_{\rm cubic}$ is a dimensionless parameter that depends on the
interactions among the different fields in the
model~\cite{Berera:2008ar,BasteroGil:2010pb}.  Hereafter, we refer to
the above $\Upsilon_{\rm cubic}$ as the {\it cubic dissipation
  coefficient}. 

The second dissipation regime we consider is obtained in a particle
physics model in which the inflaton  directly couples to the radiation
fields and gets protection from large thermal corrections due to the
symmetries obeyed by the model~\cite{Bastero-Gil:2016qru}. 
The resulting dissipative coefficient is linear in the temperature being simply given by
\begin{equation}\label{upsilonT1}
\Upsilon_{\rm linear} = C_{\rm linear} T,
\end{equation}
where here also $C_{\rm linear}$ is a dimensionless parameter that
depends on the specific interactions of the model (see, e.g.,
ref.~\cite{Bastero-Gil:2016qru} for details).  Hereafter, we refer to
the above $\Upsilon_{\rm linear}$ as the {\it linear dissipation
  coefficient}. 

The primordial power  spectrum in WI can be strongly influenced by the
presence of dissipative effects (for WI effects at the perturbation
level see  also
refs.~\cite{Graham:2009bf,BasteroGil:2011xd,Bastero-Gil:2014raa,Visinelli:2014qla})
and can be parameterized as
\begin{equation}
\label{eq:parameterization}
\Delta_\mathcal{R}\! =P_{0}(k/k_{*})\mathcal{F}(k/k_{*}),
\end{equation}
where we have defined $P_{0}(k/k_{*})\equiv(H_{*}^2/2
\pi\dot{\phi}_{*})^2$, which is the usual CI result, while
$\mathcal{F}(k/k_{*})$ corresponds to  the enhancement term  in
WI~\cite{Ramos:2013nsa} 
\begin{eqnarray} \label{spectrum}
\!\!\!\!\!\!\!\!\!\mathcal{F}(k/k_{*})\! =\!\left(\!1\!  +\!2n_*^{\rm WI}
\!+\!\frac{2\sqrt{3}\pi Q_*}{\sqrt{3\!+\!4\pi Q_*}}{T_*\over
  H_*}\!\right)\! G(Q_*).
\end{eqnarray}
where $n_*^{\rm WI}$ denotes the inflaton statistical distribution due to the presence of the radiation bath, $G(Q_*)$ accounts for the growth of inflaton fluctuations due to its coupling with the radiation fluid and the quantity $Q_*$ is the ratio
\begin{equation}
Q_*= \frac{\Upsilon(T_*,\phi_*)}{3 H_*}.
\label{Qstar}
\end{equation}

It is worth recalling again here that the term $2n_*^{\rm WI}$ in eq.~(\ref{spectrum}) appears due to the
intrinsic dissipative dynamics in WI due to particle decay and the consequent change of the vacuum state from the
Bunch-Davis one~\cite{Ramos:2013nsa}.
The third term inside the round brackets in eq.~(\ref{spectrum}) and proportional to $Q_* T_*/H_*$ appears as
a consequence that in WI the inflaton perturbations is of the form of a Langevin-like equation.  That term 
appears from the dissipation contribution to the spectrum (see also again ref.~\cite{Ramos:2013nsa} for details 
and also ref.~\cite{Hall:2003zp} for a previous derivation of this contribution). 

We note that $G(Q_*)$ can only be determined numerically by solving the full set of perturbation equations of WI~\cite{Graham:2009bf,BasteroGil:2011xd}. 
According to the method of the previous works, we use a numerical fit
for $G(Q_*)$~\cite{Benetti:2016jhf} and we consider for the linear dissipation
coefficient $\Upsilon_{\rm linear}$, that $G(Q_*)$ is given by
\begin{eqnarray} \label{growing_mode}
G_{\rm linear}(Q_*)\simeq 1+ 0.335 Q_*^{1.364}+ 0.0185Q_*^{2.315},
\end{eqnarray} 
while for the cubic dissipation coefficient,  $\Upsilon_{\rm cubic}$,
$G(Q_*)$ is given by
\begin{eqnarray} \label{growing_mode2}
G_{\rm cubic}(Q_*)\simeq 1+ 4.981 Q_*^{1.946}+ 0.127 Q_*^{4.330}.
\end{eqnarray} 
As also
considered in previous works, we here are going to assume a thermal
equilibrium distribution function $n_*^{\rm WI} \equiv n_{k_*}^{\rm WI}$ for the
inflaton such that  it assumes the Bose-Einstein distribution form,
$n_*^{\rm WI} =1/[\exp(H_*/T_*)-1]$.  The scalar spectral amplitude value at
the pivot scale is set by the CMB data as $\Delta_{{\cal R}}(k=k_*)
\simeq 2.2 \times 10^{-9}$.

Noteworthy, the eq.~(\ref{spectrum}) explicitly takes into account the fact that the dynamics in WI happens in a radiation environment. This is expressed by both the statistical distribution term $n_*^{\rm WI}$  and also by the term proportional to $Q_* T_*/H_*$. In WI by definition we have that $T > H$. Thus, when $Q<1$, which is the regime we are considering in the present work, the dominant contribution actually comes from $n_*^{\rm WI}$, which for inflaton excitations in thermal equilibrium implies that $1+2 n_*^{\rm WI} \simeq 2 T_*/H_*$. The WI contribution is then not a small correction with respect to the cold inflation case, but is in fact dominated by the thermal effects. It is because of this that in WI it is possible to make a quartic inflaton potential compatible  with the observations (see, e.g., refs. \cite{Bartrum:2013fia}, \cite{Bastero-Gil:2016qru}).

The quantities in the primordial power spectrum of
eq.~(\ref{eq:parameterization}) are then evaluated when the relevant
CMB modes  cross the Hubble radius around $N_* \approx 50 - 60$
e-folds before the end of inflation. In this work we consider
$N_*=60$ for definiteness.  

The tensor-to-scalar ratio $r$ and the spectral tilt $n_s$ in WI
follow the usual definitions, as in the CI scenario,
\begin{equation}
r= \frac{\Delta_{T}}{\Delta_{{\cal R}}},
\label{eq:r}
\end{equation}
and
\begin{equation}
n_s -1 = \lim_{k\to k_*}   \frac{d \ln \Delta_{{\cal R}}(k/k_*) }{d
  \ln(k/k_*) },
\label{eq:n}
\end{equation}
where $\Delta_{T} = 2 H_*^2/(\pi^2 M_p^2)$ is the tensor power
spectrum.  Due to the weakness of gravitational interactions, the
tensor modes are expected not to be significantly  affected by the
dissipative dynamics and $\Delta_{T}$ is unchanged compared to the CI
result~\cite{Ramos:2013nsa}.

%%%%%%%%%%%%%%%%%%%%%%%%%%%%%%%%%%%%%%%%%%%%%%%%%%%%%%%%%%%%%%%%%%%%%%%%%%
\subsection{Warm Inflation in LQC}
\label{WILQC}

In the previous subsection  we have considered the dynamics as in the
standard GR case, thus neglecting the LQC corrections to the
{}Friedmann's equation~(\ref{Hubble}).  These corrections, at the
background level, are important much before inflation sets in, when
the energy densities are very high. As we showed previously, the
consequent dynamics leads to a bounce phase, both in CI and in
WI. After the bounce, during the expansion, the energy densities
decreases such that at the onset of WI the energy densities are much
smaller than the critical density,  $\rho_* \ll \rho_{\rm cr} \simeq
258.58 \times M_{\rm Pl}^4$ and quantum effects on the geometry from
LQC can be neglected in principle.  However, although this is valid at
the background level, the same is not true at the perturbative level and the
power spectrum can receive important contributions due to LQC, as
we already discussed in the previous section. Including the correction from LQC,
eq.~(\ref{PR}), in the WI result eq.~(\ref{spectrum}),  is
equivalent to modifying the enhancement term  $\mathcal{F}(k/k_{*})$
of eq.~(\ref{spectrum}), such that
\begin{equation}
\label{eq:parameterizationLQC}
\Delta_{\mathcal{R},\rm LQC} = P_{0}(k/k_{*})\mathcal{F}_{\rm
  LQC}(k/k_{*}),
\end{equation}
where

\begin{eqnarray} \label{spectrumWILQC}
\mathcal{F}_{\rm LQC}(k/k_{*}) = \left(1 +2n_*^{\rm LQC}
+2n_*^{\rm WI} +\frac{2\sqrt{3}\pi Q_*}{\sqrt{3+4\pi Q_*}}{T_*\over
  H_*}\right) G(Q_*).
\end{eqnarray}
To fully understand why the LQC correction to the spectrum, $2n_{*}^{LQC}\equiv \delta_{PL}$, 
enters in the form as it is included in eq.~(\ref{spectrumWILQC}), i.e., additively with
the term $2n_{*}^{WI}$, it is worth recalling the origin of each of the WI terms
appearing in eq.~(\ref{spectrumWILQC}). 
In WI, the inflaton perturbations are described by a Langevin-like equation,
with both a dissipation term, $\Upsilon$, and an associated stochastic noise term, $\xi$ 
(see, for instance, refs.~\cite{Graham:2009bf,BasteroGil:2011xd,Bastero-Gil:2014jsa,Visinelli:2014qla,Ramos:2013nsa}), where they satisfy the fluctuation-dissipation theorem. 
When computing the power spectrum for the inflaton perturbations,
$\langle |\delta \phi^2| \rangle$, it will then receive two contributions, namely:
(a) the contribution from the particular solution for the $\delta \phi$ equation of motion.
This contribution depends on the two-point correlation function of the noise,
$\langle \xi \, \xi \rangle \propto Q T/H$, and it is independent of the initial conditions
(vacuum state), as it should (see, for example, ref.~\cite{Ramos:2013nsa} for an explicit derivation).
In particular it is insensitive to any details of the pre-inflationary phase and the bounce.
As such, we do no expect any correction factor due to the bounce  multiplying this term
or any other of those details related to the initial conditions affecting this term;
(b) then there is the contribution to $\langle |\delta \phi|^2 \rangle$ that comes from the
homogeneous part of the equation of motion for $\delta \phi$. This contribution does depend explicitly
on the initial conditions, i.e., the vacuum state. But it happens that in WI,
because of the dissipation, particle production and the existence of a radiation bath,
this initial state for the inflaton perturbations is not simply the Bunch-Davis vacuum,
but it is an {\it excited state}. This gives the origin of the term $2n_*^{\rm WI}$ in eqs.~(\ref{spectrum}) 
and (\ref{spectrumWILQC}). 
But the LQC factor  $\delta_{PL} \equiv 2 n^{\rm LQC}$ is itself also
a change of the Bunch-Davis  vacuum due to the gravitational particle production due to the
bounce and which is carried over till the moment the relevant scales leave the Hubble radius.
Since both contributions $2 n^{\rm WI}$ and $2 n^{\rm LQC}$ express the change of
the Bunch-Davis vacuum due to particle production, hence, $2 n^{\rm WI} + 2 n^{\rm LQC}$ 
describes this total effect on
the spectrum due to the different sources of particle production.
{}Finally, for completeness, there is the factor $G(Q)$ in eqs.~(\ref{spectrum}) 
and (\ref{spectrumWILQC}). As already explained,
this factor appears as a correction to the spectrum due to the coupling of the inflaton and radiation
perturbations (see, e.g., refs.~\cite{Graham:2009bf,BasteroGil:2011xd}). 
As we do not have an explicit analytical expression
for this effect, $G(Q)$ is obtained by fitting the full numerical result for the perturbations spectrum
(see refs.~\cite{Graham:2009bf,BasteroGil:2011xd,Bastero-Gil:2014jsa}).
Since the behavior of the spectrum with $Q$ is smooth and well
behaved, we can always do this procedure (numerical fitting) with a sufficient
precision such that any arbitrariness in the fitting function does not change the observable
quantities, e.g., $n_s$ and $r$.

{}Finally, as far the tensor spectrum is concerned, one notes that
the correction term $2n_*^{\rm LQC} \equiv \delta_{PL}$ 
affects
both the scalar and tensor perturbations, i.e., the GR result for the
tensor perturbations $\Delta_{T} = 2 H_*^2/(\pi^2 M_p^2)$, gets now
modified according to eq.~(\ref{PH}).  Although this modification does not lead to
a change in the tensor-to-scalar ratio value in the CI in the LQC
case, it change $r$ in the WI in the LQC case since the factor
$\left(1 +\delta_{PL}\right)$ in eq.~(\ref{PH}) does not simplify with
the one appearing in the scalar case through eq.~(\ref{eq:parameterizationLQC}).  It is true that the LQC correction factor would cancel in the tensor-to-scalar ratio $r$  also in WI provided that the thermal effects would be small in the latter, which is not the case of the models we consider here.

As an additional note, let us comment on the validity of the use of the result (\ref{delta}) also in the
WI case. Since WI is in principle characterized by significant radiation production,
we should be concerned whether this radiation might also extend way back to the quantum bounce time
and, thus, potentially affecting the derivation of the quantum bounce effect, expressed by the contribution in 
eq.~(\ref{delta}).
{}For this, it is useful to first give an estimate for the radiation energy density $\rho_R$ in the WI regime
and also the amount of energy density in the form of kinetic energy for the inflaton field, $\dot \phi^2/2$.
{}From the background equations (\ref{rho}), (\ref{Vphi}) and (\ref{eqrhoR}), we find for instance that during the warm inflationary slow-roll phase,
\begin{eqnarray}
\frac{\rho_R}{\rho} &\simeq& \frac{\epsilon_H}{2} \frac{Q}{1+Q},
\label{rhoRrho}
\\
\frac{\dot \phi^2/2}{\rho} &\simeq& \frac{\epsilon_H}{3} \frac{1}{1+Q},
\label{kineticrho}
\end{eqnarray}
where $\epsilon_H$ is the Hubble slow-roll parameter,
\begin{equation}
\epsilon_H \equiv -\frac{\dot H}{H^2}.
\end{equation}
Since during inflation we have by definition $\epsilon_H \ll 1$ and $\rho \approx V(\phi)$, 
the eqs.~(\ref{rhoRrho}) and (\ref{kineticrho}) show that the energy density in the form
of the inflaton potential dominates both over the radiation energy density 
and the kinetic energy density of the inflaton field during the slow-roll phase. {}Furthermore, in the weak dissipative
regime of WI, for which $Q<1$, the kinetic energy density of the inflaton tends to dominate over the
radiation one. This is particularly true for the two dissipation models we analyze 
in this work, where both lead to consistent results with the Planck measurements
mostly when we are in the weak dissipative regime. Since the bounce is typically dominated
by the kinetic energy density, this should continue to be true if we 
extend the dynamics in the WI case way back to the bounce in LQC. Even in the large dissipative regime,
$Q >1$, where we can have the quantity in (\ref{rhoRrho}) to be larger than the one in (\ref{kineticrho})  during the slow-roll phase, 
 we recall that whenever the bounce is dominated by the kinetic energy, the immediate
dynamics of the universe is like that of stiff matter, i.e., the dominant energy density
behaves like $\propto 1/a^6$. Thus, even though the kinetic energy density might be smaller
during inflation, it should be larger than radiation anyway by the bounce time.
Even though the question of how the presence of a non-negligible radiation at the bounce time
might affect the quantity $\delta_{PL}$ is an important one, addressing this issue in more detail is
beyond the scope of the present work. Thus, we will always be assuming that the assumptions
used by the authors of ref.~\cite{zhu} to derive the result eq.~(\ref{delta}) will continue 
to be valid here, in particular that
the bounce is always dominated by the kinetic energy of the inflaton field.

%%%%%%%%%%%%%%%%%%%%%%%%%%%%%%%%%%%%%%%%%%%%%%%%%%%%%%%%%%%%%%%%%%%%%%%%%%
\section{Analysis and results}
\label{analysis}

Let us describe in this section our strategy used in the analysis and
the results we obtain following it.

%%%%%%%%%%%%%%%%%%%%%%%%%%%%%%%%%%%%%%%%%%%%%%%%%%%%%%%%%%%%%%%%%%%%%%%%%%
\subsection{Strategy for the analysis}

To perform our analysis, we consider a minimal $\Lambda$CDM model and
modify the standard primordial power-law spectra following the above
equations for the WI in LQC, i.e, parameterizing the scalar primordial
spectrum as eq.~(\ref{eq:parameterizationLQC}), and likewise dealing
with the tensor spectrum, eq.~(\ref{PH}).  Therefore, we vary the
usual cosmological parameters, namely, the physical baryon density,
$\Omega_{b} h^{2}$, the physical cold dark matter density, $\Omega_{c}
h^{2}$, the ratio between the sound horizon and the angular diameter
distance at decoupling, $\theta$, and the optical depth, $\tau$.  In
addition, we have one more parameter, $k_{B}/a_{0}$ \footnote{We note
  that $a_{0}=1$, so hereafter $k_{B}/a_{0}=k_{B}$.}, which is related
to the effects of  the pre-inflationary dynamics due to LQC and that
appears explicitly in the expression for $\delta_{PL}$,
eq.~(\ref{delta}). 

Noteworthy, when we analyze the WI scenario, we do not use the
primordial parameters $A_{s}$, $n_{s}$ and $r$, respectively the
scalar amplitude, the spectral index and the tensor-to-scalar-ratio,
as free parameters in our analysis. Both $P_{0}(k/k_{*})$ and
$\mathcal{F}(k/k_{*})$ of eq.~(\ref{eq:parameterizationLQC}) (and
similarly for the tensor case) are obtained numerically for the
studied models by solving the background equations for the WI in LQC
(for simplicity, we refer to the model hereafter as WI+LQC) for
different values of the the dissipation ratio $Q_{*}$~\cite{Benetti:2016jhf}. 
These values are calculated for the scales leaving the Hubble radius in an interval
$\Delta N = 5$ around the value of  $N =60$ for which we assume that
the pivot scale crosses the horizon, and $P_{0}(k/k_{*})$ is
normalized to the amplitude value of the standard $\Lambda$CDM
model~\cite{Ade:2015lrj} in each model we consider~\footnote{Our strategy 
is similar to what was used previously in ref.~\cite{Benetti:2016jhf}. We 
stress that different strategies were adopted later by the authors of 
refs.~\cite{Bastero-Gil:2017wwl,Arya:2017zlb}, but both obtained results
analogous to the ones found in ref.~\cite{Benetti:2016jhf}.}.

We note that the dissipation ratio $Q_{*}$, the temperature ratio
$T_{*}/H_{*}$ and the amplitude $P_{0}(k/k_{*})$ of
eq.~(\ref{eq:parameterizationLQC}) are of power-law form with the scale for the 
considered potential in both the dissipation regimes we studied here. Hence, we can
approximate them in our analysis with a power-law fitting without loss
of  information.  In our analysis we also vary the nuisance foreground
parameters~\cite{Aghanim:2015xee} and consider purely adiabatic
initial conditions. The sum of neutrino masses is fixed to $0.06$ eV
and we consider for the pivot $k_{*} = 0.002$ Mpc$^{-1}$. Also, we
work with flat priors for the cosmological parameters, and assume a
flat prior for the $k_{B}$ parameter varying in the range $[0:0.1]$
(in units of Mpc$^{-1}$).

We perform a primary analysis with Mathematica~\cite{Mathematica},
obtaining the required parameterizations of eqs.~(\ref{PH}) and
(\ref{eq:parameterizationLQC}), for different values of the
dissipation ratio $Q_{*}$.  Then, we use a modified version of the
CAMB~\cite{CAMB} to compute the theoretical CMB anisotropies spectrum
in the WI+LQC context using such parameterizations and then employ a Monte
Carlo Markov Chain analysis via the publicly available package
CosmoMC~\cite{CosmoMC} in order to compare these theoretical
predictions with observational data. We choose to use the latest release of 
Planck data (2015) at both low and high multipoles~\cite{Aghanim:2015xee} 
(hereafter CMB), considering also the B-mode polarization data from the BICEP2
Collaboration~\cite{3,4} to constrain the parameters associated with
the tensor spectrum, using the combined BICEP2/Keck-Planck likelihood
(hereafter BKP).
 
%%%%%%%%%%%%%%%%%%%%%%%%%%%%%%%%%%%%%%%%%%%%%%%%%%%%%%%%%%%%%
 \begin{table}[t]
\centering
\caption{
  Confidence limits for the cosmological parameters in the Cold Inflation
  in LQC model, using CMB+BKP data.}
\label{tab:1}
\begin{tabular}{cc}
\hline 
\hline
{ Parameters }& 
{Cold Inflation in LQC}\\

\hline
$\Omega_{b} h^{2}$  & $0.02218 \pm 0.00023$ \\

$\Omega_{c} h^{2}$  & $0.1203 \pm 0.0021$ \\

$100 \, \theta$  & $1.04081 \pm 0.00048$ \\

$\tau$  & $0.075 \pm 0.015$ \\
 
${\rm{ln}}(10^{10} A_s)$ & $3.199 \pm 0.031$ \\
 
$ns$ & $0.9648 \pm 0.0061$ \\
 
$r$ & $ < 0.027$ in 1$\sigma$ ($ < 0.054$ in 2$\sigma$) \\
 
$k_{B}$ (Mpc$^{-1}$) & $ < 1.9\times 10^{-4}$ in 1$\sigma$ ($< 3.3\times 10^{-4}$ in 2$\sigma$)\\
\hline 
\hline
\end{tabular}
\end{table} 
%%%%%%%%%%%%%%%%%%%%%%%%%%%%%%%%%%%%%%%%%%%%%%%%%%%%%%%%%%%%%
%%%%%%%%%%%%%%%%%%%%%%%%%%%%%%%%%%%%%%%%%%%%%%%%%%%%%%%%%%%%%
\subsection{Results}
\label{results}

Let us start analyzing the case of CI in LQC. In this case we use the
eqs.~(\ref{PR}) and (\ref{PH}) in the CAMB code, also considering the
standard free parameters $A_{s}$, $n_{s}$ and $r$ in our analysis. Our
results are shown in table~\ref{tab:1} and, starting from the
upper value of $k_{B}$, we can calculate the required extra number of e-folds 
in CI + LQC (see eq.~(\ref{kbN})). We obtain $\delta N \gtrsim 21$ at $1\sigma$, 
or $\delta N \gtrsim 20.4$ in $2\sigma$. These values are found to be in good 
agreement with previous results obtained by the authors of ref.~\cite{zhu, 14pp}. 

{}Focusing now at the WI+LQC model, we use the
parameterizations discussed in the previous subsection. Hence, we make
an accurate analysis by selecting several models with different $Q_*$
values, then perform an Monte Carlo Markov chain (MCMC) analysis in order to constrain the $k_B$ value.  
Let us stress again that in this case we do not use the standard
free parameters $n_{s}$ and $r$ in our analysis, which instead are
explicitly computed and they are
fixed by the chosen value of $Q_*$ as reported in table~\ref{tab:2},
where we show the results of our analysis for both cases of linear and
cubic dissipation regimes for the selected models. 
%
%%%%%%%%%%%%%%%%%%%%%%%%%%%%%%%%%%%%%%%%%%%%%%%%%%%%%%%%%%%%%
\begin{table}[t]
\caption{
  Upper limits on the parameter $k_B$ in 1$\sigma$ for the models
  analyzed in the linear and cubic dissipation regimes.
  The 2$\sigma$ values are also shown in the square brackets.}
\label{tab:2}
\newcommand{\mc}[3]{\multicolumn{#1}{#2}{#3}}
\begin{center}
\begin{tabular}{cccc}
\hline
\hline
\mc{4}{c}{Linear Dissipation}\\
\hline
$Q_*$ & \mc{1}{c}{$r$} & \mc{1}{c}{$n_s$} & \mc{1}{c}{$k_B$ (Mpc$^{-1}$) [2$\sigma$]}\\
\hline
$4\times 10^{-6}$ & \mc{1}{c}{$0.050$ } & \mc{1}{c}{0.9655} & \mc{1}{c}{$< 0.00038\;$ [ $<0.00071$ ]}\\
$3\times 10^{-4}$ & \mc{1}{c}{$0.012$} & \mc{1}{c}{0.9659} & \mc{1}{c}{$<0.00072\;$ [ $<0.00138$ ]}\\
$5\times 10^{-2}$ & \mc{1}{c}{$0.002$} & \mc{1}{c}{0.9632} & \mc{1}{c}{$<0.00167\;$ [ $<0.00328$ ]}\\
$1.02$ & \mc{1}{c}{$1.7\times 10^{-4}$} & \mc{1}{c}{0.9663} & \mc{1}{c}{$<0.00526\;$ [ $<0.00973$ ]}\\
$1.65$ & \mc{1}{c}{$6.8\times 10^{-5}$} & \mc{1}{c}{0.9722} & \mc{1}{c}{$<0.00675\;$ [ $<0.01318$ ]}\\
$2.08$  & \mc{1}{c}{$4.1\times 10^{-5}$} & \mc{1}{c}{0.9756} & \mc{1}{c}{$<0.00801\;$ [ $<0.01541$ ]}\\
$3.19$ & \mc{1}{c}{$1.35\times 10^{-5}$} & \mc{1}{c}{0.9825} & \mc{1}{c}{$<0.00901\;$ [ $<0.01873$ ]}\\
$4.38$ & \mc{1}{c}{$5.3\times 10^{-6}$} & \mc{1}{c}{0.9874} & \mc{1}{c}{$<0.01112\;$ [ $<0.02267$ ]}\\
\hline
\hline
\mc{4}{c}{Cubic Dissipation}\\
\hline
$Q_*$ & \mc{1}{c}{$r$} & \mc{1}{c}{$n_s$} & \mc{1}{c}{$k_B$ (Mpc$^{-1}$) [2$\sigma$]}\\
\hline
$8\times 10^{-5}$ & \mc{1}{c}{$0.070$ } & \mc{1}{c}{0.9742} & \mc{1}{c}{$<0.00035\;$ [ $<0.00068$ ]}\\
$1.8\times 10^{-4}$ & \mc{1}{c}{$0.058$} & \mc{1}{c}{0.9734} & \mc{1}{c}{$<0.00042\;$ [ $<0.00080$ ]}\\
$4.4\times 10^{-4}$ & \mc{1}{c}{$0.046$} & \mc{1}{c}{0.9720} & \mc{1}{c}{$<0.0044\;$ [ $<0.00086$ ]}\\
$8\times 10^{-4}$ & \mc{1}{c}{$0.040$} & \mc{1}{c}{0.9709} & \mc{1}{c}{$<0.0049\;$ [ $<0.00094$ ]}\\
$2 \times 10^{-3}$ & \mc{1}{c}{$0.030$} & \mc{1}{c}{0.9687} & \mc{1}{c}{$<0.0062\;$ [ $<0.00116$ ]}\\
$4\times 10^{-3}$  & \mc{1}{c}{$0.026$} & \mc{1}{c}{0.9680} & \mc{1}{c}{$<0.0063\;$ [ $<0.00122$ ]}\\
$6 \times 10^{-3}$ & \mc{1}{c}{$0.023$} & \mc{1}{c}{0.9677} & \mc{1}{c}{$<0.0066\;$ [ $<0.00130$ ]}\\
$ 3.3 \times 10^{-2}$ & \mc{1}{c}{$0.014$} & \mc{1}{c}{0.9747} & \mc{1}{c}{$<0.0092\;$ [ $<0.00174$ ]}\\
\hline
\hline
\end{tabular}
\end{center}
\end{table}
%%%%%%%%%%%%%%%%%%%%%%%%%%%%%%%%%%%%%%%%%%%%%%%%%%%%%%%%%%%%%%%%%%%%%%%%%%

\begin{center}
\begin{figure}[t]
\subfigure[ Linear dissipation case.]{\includegraphics[width=7.2cm]{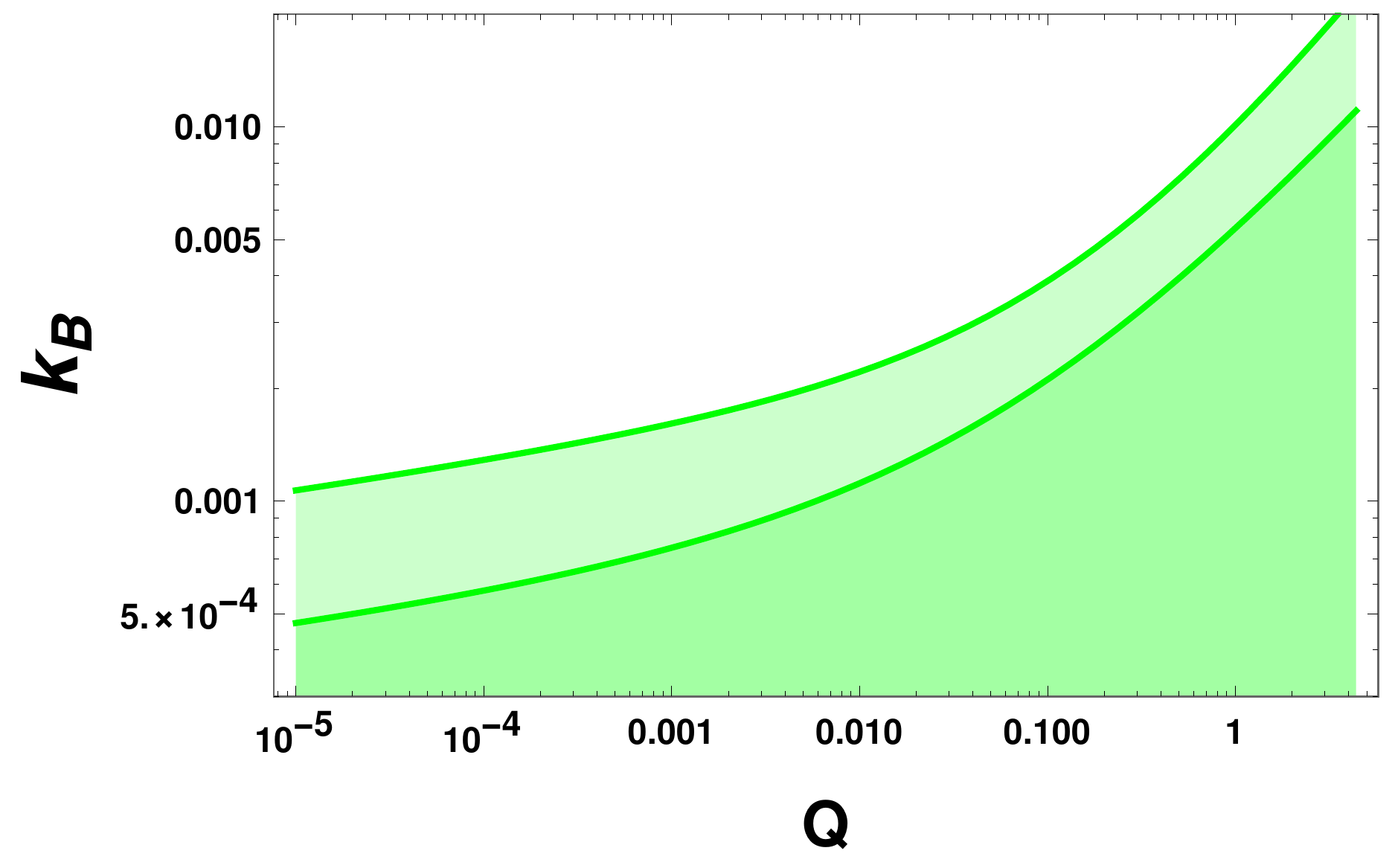}}
\subfigure[ Cubic dissipation case.]{\includegraphics[width=7.2cm]{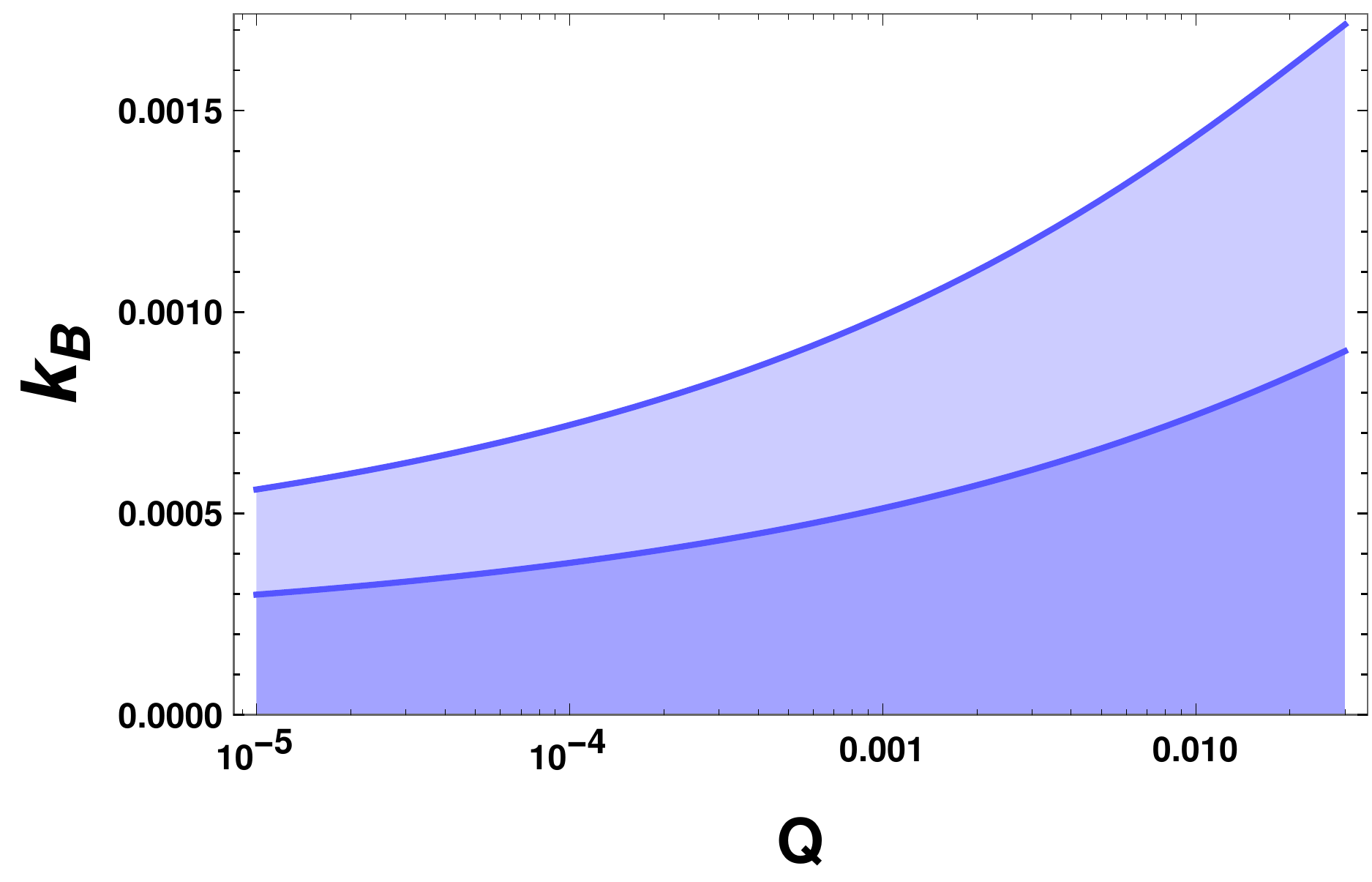}}
\caption{$k_B$ confidence bounds at 1-$\sigma$ (dark colour) and 2-$\sigma$ (light colour).}
\label{fig2}
\end{figure}
\end{center}

{}For simplicity, we do not report in the table~\ref{tab:2} the values of the other cosmological parameters since they are in fully agreement
with the ones of the standard model~\cite{Ade:2015lrj}.  We can note a
very striking behavior of $k_{B}$ with $Q_*$, i.e., the upper limit of
$k_{B}$ increases with $Q_*$ in both the considered regimes, with more
pronounced growth in the linear dissipative regime. This is also
illustrated in {}fig.~\ref{fig2}, where we show the 1-$\sigma$ and
2-$\sigma$ regions for these parameters through the shaded areas for both 
the dissipation cases studied here. 
 
Let we stress that for $Q_* \rightarrow 0$ the value of $k_{B}$ tends
to the one obtained in CI, as expected.  
{}Furthermore, the behavior of increasing values of $k_{B}$ with $Q_*$ is also 
foregone for other forms for the inflaton potential or other
forms for the dissipation coefficient in general. This is expected
because the presence of dissipation always tend to damp oscillatory
features in the spectrum, thus pushing the bound on $k_B$ to larger
values.  Larger allowed values of $k_B$ in this case are potentially
important in the LQC case, by allowing the bounce to happen closer to
the point $N_*$, where the physical scales crossed the Hubble radius,
in the universe evolution. 

We can now infer for the WI+LQC models studied above which are the extra number
of e-folds required, from the bounce instant till the moment the physical
scales crossed the Hubble radius during inflation at $N_*$, using the
values of $k_{B}$ constrained in our analysis and the relation given in
eq.~(\ref{kbN}).  {}For the linear dissipative case, we obtain $\delta
N\geq 17$ e-folds in $1\sigma$ ($\delta N\geq 16$ in $2\sigma$) for
the model with the highest value of $Q_*$ analyzed ($Q_*=4.38$). While we obtain 
$\delta N\geq 20$ ($\delta N\geq 19.7$ in $2 \sigma$) for the model
with lowest value of $Q_*$ ($Q_*=4\times 10^{-6}$). These values are
similar to the ones obtained in the cubic dissipative case, where we find
$\delta N\geq 19$ e-folds in $1\sigma$ ($\delta N\geq 18$ in $2
\sigma$) for $Q_*=3.3\times10^{-2}$ and $\delta N\geq 20$ ($\delta
N\geq 19.7$ in $2 \sigma$) for $Q_*=8\times 10^{-5}$.  We can note
that the higher dissipation values require the least extra number
of e-folds from the bounce till $N_*$.  As a consequence, the LQC
bounce in the WI case can happen relatively later (closer to $N_*$)
than in the CI + LQC
case, and still allowing the modifications in the spectra (scalar and
tensorial) of perturbations to be  consistent with observations. 
%
%%%%%%%%%%%%%%%%%FIGURE01%%%%%%%%%%%%%%%%%%%
%%%%%%%%%%%%%%%%%%%%%%%%%%%%%%%%%%%%%%%%
%
%%%%%%%%%%%%%%%%%FIGURE02%%%%%%%%%%%%%%%%%%%
\begin{center}
\begin{figure}[t]
\subfigure[ Linear dissipation
  case.]{\includegraphics[width=7.2cm]{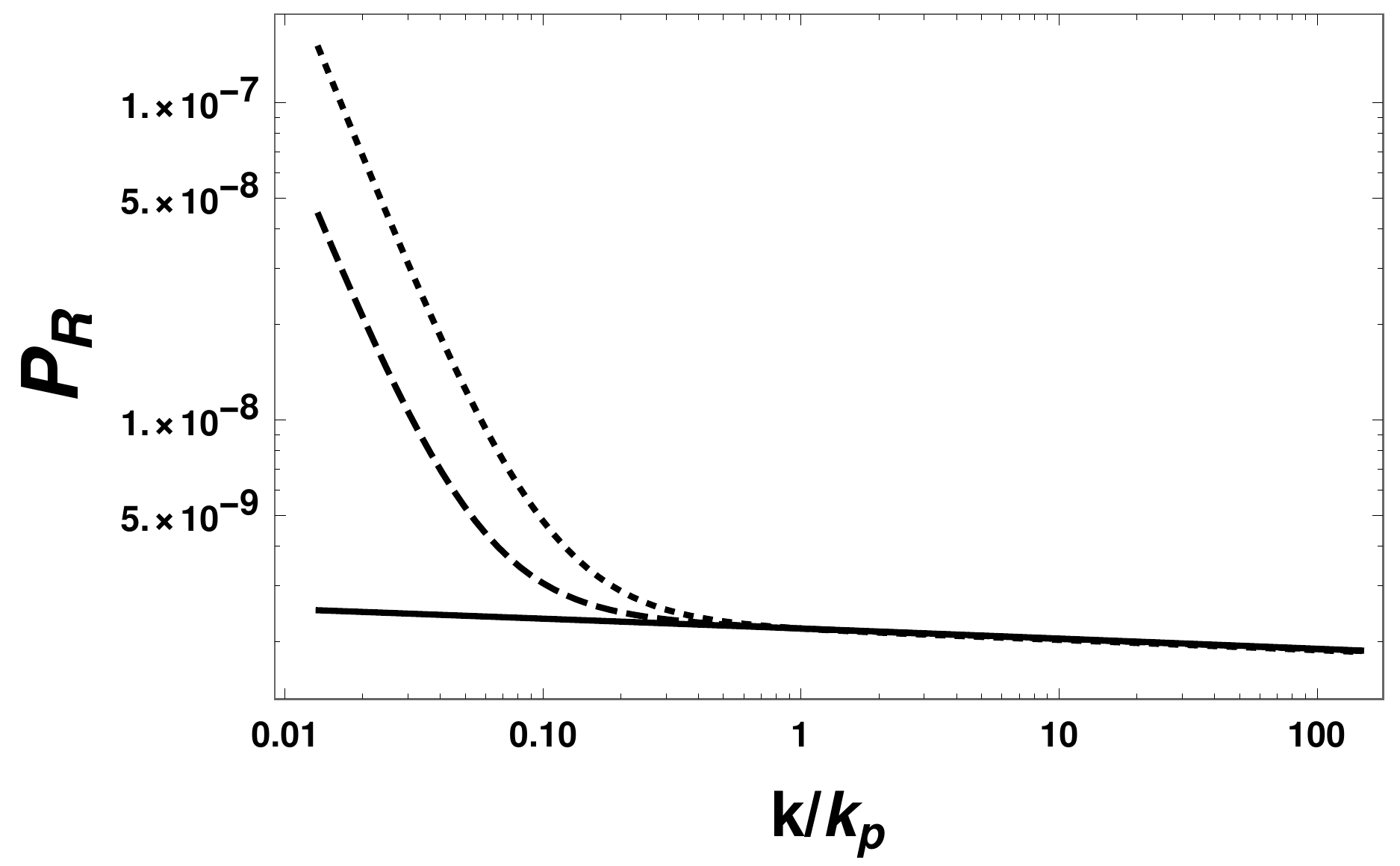}}
\subfigure[ Cubic dissipation
  case.]{\includegraphics[width=7.2cm]{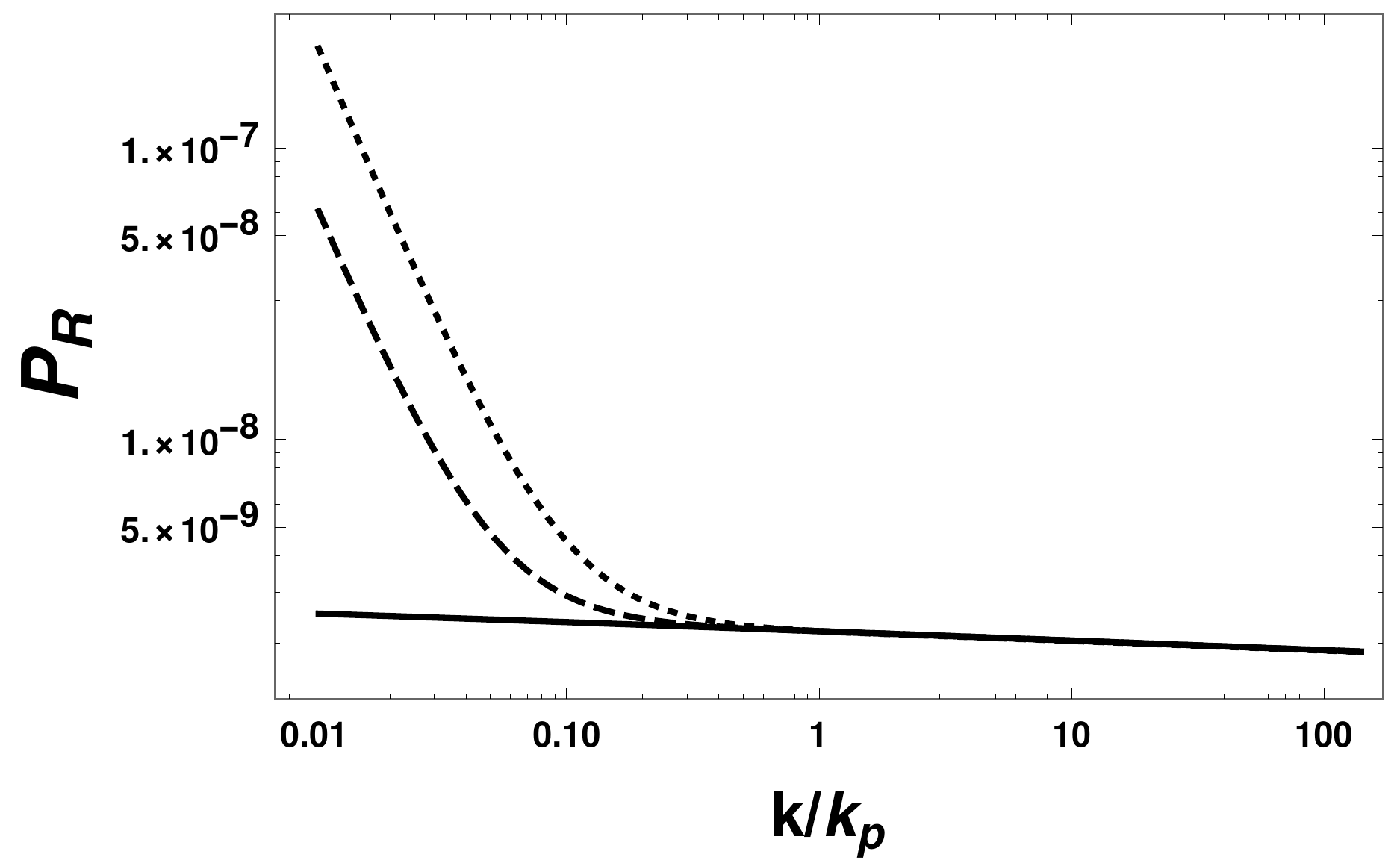}}
\caption{The amplitude of the spectrum as a function of $k/k_{p}$,
  where $k_{p}$ is the pivot scale, $k_{p}=0.002$ Mpc$^{-1}$. The
  linear dissipative regime is for the case with $Q_*=1.02$, while the
  cubic dissipative regime is for $Q_*=0.006$. The solid line
  corresponds to the WI model without the LQC correction
  ($\delta_{PL}=0$), the dashed line corresponds to the case including
  LQC corrections for the $k_{B}$ maximum  allowed in this model at
  1-$\sigma$ level. The dotted line for the $k_{B}$ maximum  allowed
  at 2-$\sigma$ level.}
\label{fig:spectraWI}
\end{figure}
\end{center}
%%%%%%%%%%%%%%%%%%%%%%%%%%%%%%%%%%%%%%%%
%
In order to further see the effects of the WI+LQC in the observables,
we show the primordial power spectrum of WI+ LQC for two particular
models considered in table~\ref{tab:2}. In fig.~\ref{fig:spectraWI} we
have thus chosen to use the model with $Q_*=1.02$ for the linear
dissipative regime, shown in the panel (a), and the case with $Q_*=0.006$ 
for the cubic dissipative regime, shown in the panel (b). 
We can note that WI with LQC
correction (dashed and dotted lines, respectively, obtained using the
upper limit values of $k_{B}$ at 1-$\sigma$ and at 2-$\sigma$)
increases the power at lower values of $k$, i.e., for large scales,
with respect to the simplest WI model (solid line).  This behavior is
also clear in fig.~\ref{fig:TT}, where we show the temperature
anisotropy power spectrum of CMB for the the best fit model values, for
the same cases as before,  
with respect to the CI+LQC model. {}For the $k_B$ parameter, we use the values obtained 
at 1-$\sigma$ and at 2-$\sigma$ that were reported in table~\ref{tab:1} and
table~\ref{tab:2}, for the CI+LQC and WI+LQC cases, respectively. 
We can note the lower power at low multipoles of the WI+LQC models with respect
to the CI+LQC case \footnote{The relatively less power seen in fig. 3 for small multipoles (and for the $1\sigma$ values for $k_B$) can be attributed to the fact that $Q$ is not a constant in WI and it always grows with the number of e-folds, or equivalently, with time, which ends up reflecting in the anisotropy power spectrum as a function of multipoles  to have a smaller power than in the cold inflation case. This happens  since the dissipation in WI affects different scales in a different intensity.}.  
The LQC correction to the primordial spectrum thus tends to produce more power 
at low multipoles of CMB the larger is $k_B$ \footnote{ Note that 
the dominant contribution here of smaller values of $k$ having more power is entirely due to the LQC effects.
The point is that in WI we are allowed to have a larger $k_B$ with respect to the case of cold inflation, but on the other hand this injects more power at lower $k$. There is of course a limit on how much we can increase the dissipation, which is constrained by the observations, and this in fact translates in how much we can increase
$k_B$ by increasing the dissipative effects from WI. However in the models we analyzed we did not considered such higher values of $Q$ and $k_{B}$.}.
Taking into account the lower sensitivity of the data in such a
region,  the differences between the spectra using the best fit values
do not lead to a significant difference in the $\chi^2$ values, they
are about the same in all the cases considered and, thus, we are not
showing them explicitly here.  Hence, despite the non-trivial
modifications in the power spectrum due to the  presence of
dissipation in addition to the presence of a pre-inflationary dynamics
from  LQC, our analysis shows that the WI model can explain the
current observables also in the context of LQC.  At the same time, we can see 
bounds on the $k_B$ value from the LQC correction to the spectra  that allow 
for a larger variation (or freedom)  than in the case of CI+LQC.

%%%%%%%%%%%%%%%%%FIGURE03%%%%%%%%%%%%%%%%%%%
\begin{figure}[t]
	\centering \includegraphics[width=0.5\textwidth,
          angle=-90]{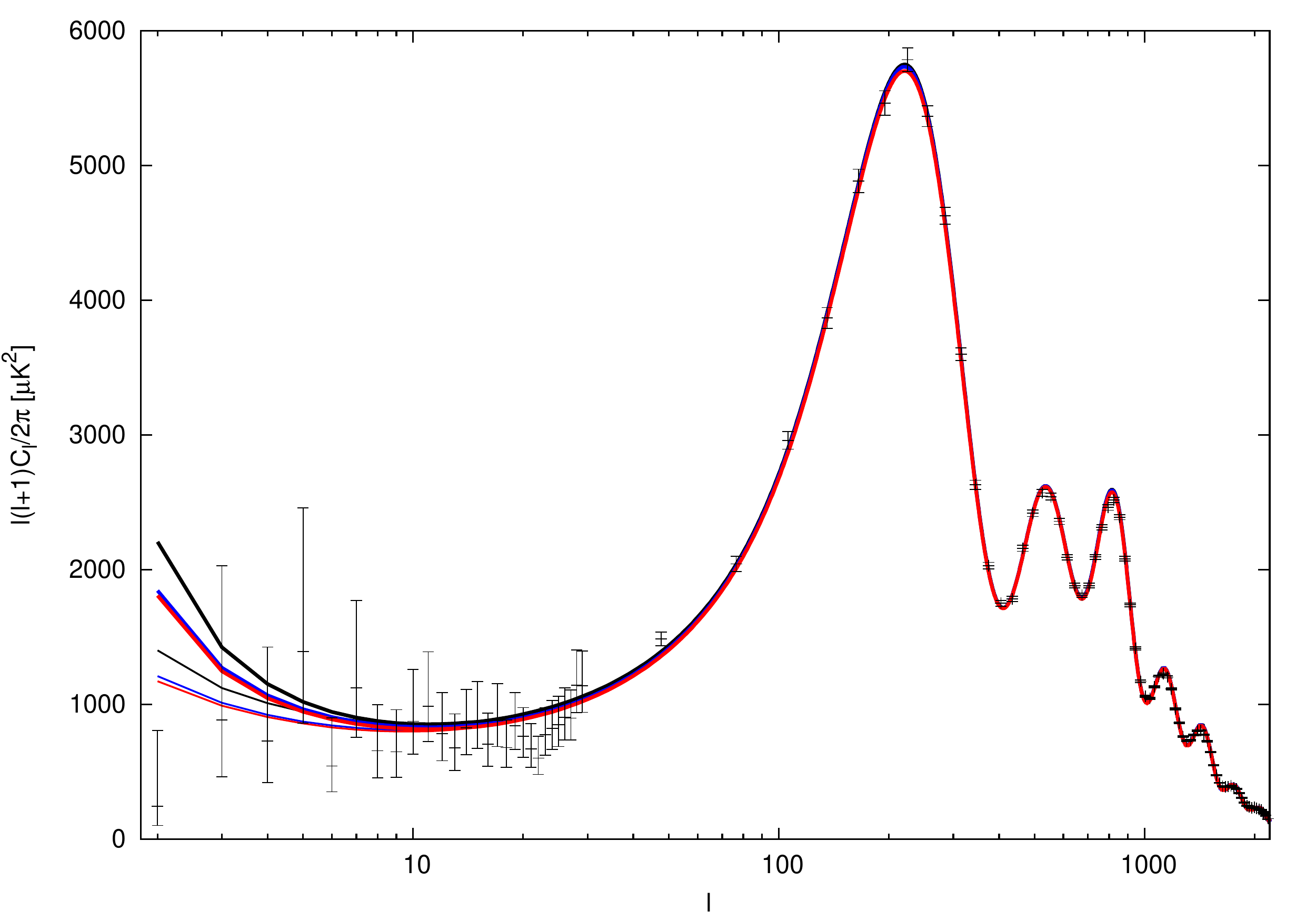}
				\caption{CMB temperature anisotropy
                                  power spectrum for the best fit
                                  models of WI+LQC in the linear
                                  dissipation regime with $Q_*=1.02$
                                  (blue lines) and for the cubic
                                  dissipative regime with $Q_*=0.006$
                                  (red lines), with respect to the
                                  CI+LQC model (black line). The
                                  models where it was assumed the
                                  $k_B$ parameter in its 1-$\sigma$
                                  values are draw with the thinner
                                  lines, while the thicker ones refer
                                  to 2-$\sigma$ values.}
	\label{fig:TT}
\end{figure}
%%%%%%%%%%%%%%%%%%%%%%%%%%%%%%%%%%%%%%%%%%%%%%%%%%%%%%%
%
%%%%%%%%%%%%%%%%%%%%%%%%%%%%%%%%%%%%%%%%%%%%%%%%%%%%%%%%%%%%%%%%%%%%%%%%%%
\section{Conclusions}
\label{conclusion}

In this work we have considered the warm inflationary scenario in the
context of LQC. The modifications in the standard spectrum due to the
pre-inflationary dynamics of LQC and also due to the presence of
dissipation during inflation, were examined in the light of the legacy
Planck data (2015). 

A noteworthy result we get in our analysis is that the upper limit in the 
value of the LQC parameter scale $k_{B}$ increases with the value of the 
dissipative ratio $Q_*$ in both dissipative regimes
that we have analyzed in this paper. Let us recall that models with
higher values of $Q_*$ ($Q_*\gg 1$) have the potential advantage  of
allowing sub-Planckian initial values for the inflaton field excursion
in the WI scenario~\cite{Motaharfar:2018zyb}.  Correspondingly, this
would lead to models which allow  higher values of the LQC parameter
scale $k_B$, pushing the bounce point closer to the point $N_*$
where the relevant physical scales crossed the Hubble radius in the
universe history. By making the quantum bounce happen closer to $N_*$, it can
potentially make it be observable through future CMB precision
measurements. 

We found that in WI, the bounce can happen at least $\delta N_* \geq 17$
e-folds before $N_*$ and the modifications in the perturbation spectra
still to be consistent with the recent observations. This result can
be compared with the one obtained for CI in LQC, $\delta N_{*} \geq
21$.  An additional result obtained from our analysis is that
models with higher dissipation requires smaller $\delta N$
values. Therefore,  WI in LQC requires less extra number of e-folds
than CI in LQC. Since it has been suggested in the
literature~\cite{Motaharfar:2018zyb} that models with higher values of
dissipation can be viable (see also ref.~\cite{Bastero-Gil:2019gao} for a recent
explicit realization of such a WI scenario), this opens the possibility for a much
closer beginning for the quantum bounce in these WI models. 

%%%%%%%%%%%%%%%%%%%%%%%%%%%%%%%%%%%%%%%%%%%%%%%%%%%%%%
\section{Acknowledgements}

 M.B. acknowledge INFN Sez. di Napoli (Iniziativa Specifica
QGSKY) for financial support. L.L.G. acknowledge financial support of the {}Funda\c{c}\~ao Carlos
Chagas Filho de Amparo \'a Pesquisa do Estado do Rio de Janeiro
(FAPERJ).  R.O.R. is partially supported by
research grants from Conselho Nacional de Desenvolvimento
Cient\'{\i}fico e Tecnol\'ogico (CNPq), Grant No. 302545/2017-4, and
Funda\c{c}\~ao Carlos Chagas Filho de Amparo \`a Pesquisa do Estado do
Rio de Janeiro (FAPERJ), Grant No. E-26/202.892/2017.   We
acknowledge the use of the High Performance Computing Center at the
Universidade Federal do Rio Grande do Norte - NPAD/UFRN - for
providing the computational facilities to run our analysis and also
the National Observatory of Rio de Janeiro (ON) for the computational support.

%%%%%%%%%%%%%%%%%%%%%%%%%%%%%%%%%%%%%%%%%%%%%%%%%%%%%%%%%%%%%

\end{document}